\documentclass[a4paper]{jpconf}
\usepackage{graphicx}
\usepackage{amssymb,amsmath,stmaryrd}

\catcode`@=11
\renewcommand{\section}{\@startsection{section}{1}{0pt}{\medskipamount}
{\medskipamount}{\bf}}
\numberwithin{equation}{section}
\catcode`@=12

\def\a{\alpha}
\def\b{\beta}
\def\g{\gamma}
\def\de{\delta}

\def\ve{\varepsilon}
\def\z{\zeta}
\def\h{\eta}
\def\th{\theta}

\def\m{\mu}
\def\n{\nu}
\def\r{\rho}
\def\s{\sigma}
\def\p{\phi}
\def\vp{\varphi}
\def\j{\psi}

\def\Ga{\Gamma}
\def\La{\Lambda}
\def\Th{\Theta}
\def\Ups{\Upsilon}

\def\da{\dot\alpha}
\def\db{\dot\beta}

\def\1{\dot 1}
\def\2{\dot 2}

\def\tU{\textrm{U}}

\def\sfrac#1#2{{\textstyle\frac{#1}{#2}}}

\def\pa{\partial}

\def\sta{\star}
\def\>{\rangle}
\def\<{\langle}
\def\+{\dagger}
\def\={\ =\ }

\newcommand{\C}{\mathbb C}
\newcommand{\R}{\mathbb R}

\newcommand{\NN}{{\mathbb{N}}}
\newcommand{\unity}{\boldsymbol{1}}

\newcommand{\Hcal}{{\cal H}}
\newcommand{\Acal}{{\cal A}}
\newcommand{\Bcal}{{\cal B}}
\newcommand{\Ccal}{{\cal C}}
\newcommand{\Ncal}{{\cal N}}
\newcommand{\Fcal}{{\cal F}}

\newcommand{\im}{\mathrm{i}}
\newcommand{\diff}{\mathrm{d}}
\newcommand{\vel}{\mathrm{v}}

\newcommand{\be}{\begin{equation}}
\newcommand{\ee}{\end{equation}}
\newcommand{\bea}{\begin{eqnarray}}
\newcommand{\eea}{\end{eqnarray}}
\newcommand{\bal}{\begin{aligned}}
\newcommand{\eal}{\end{aligned}}
\newcommand{\non}{\nonumber}
\newcommand{\und}{\qquad{\text{and}}\qquad}
\begin{document}
\title{Supersymmetric noncommutative solitons~\footnote{
Talk given during the conferences 
``Noncommutative Spacetime Geometries'' at Alessandria, March 2007, and 
``Noncommutative Geometry and Physics'' at Orsay, April 2007}}

\author{Olaf Lechtenfeld}

\address{Institut f\"ur Theoretische Physik, Leibniz Universit\"at Hannover\\
Appelstra\ss{}e 2, 30167 Hannover, Germany}

\ead{lechtenf@itp.uni-hannover.de}

\begin{abstract}
I consider a supersymmetric Bogomolny-type model in 2+1 dimensions 
originating from topological string theory. By a gauge fixing this model 
is reduced to a supersymmetric U($n$) chiral model with 
a Wess-Zumino-Witten-type term in 2+1 dimensions. After a noncommutative 
extension of the model, I employ the dressing method to construct explicit 
multi-soliton configurations on noncommutative $\R^{2,1|2\Ncal}$.
\end{abstract}

\section{\bf Introduction} \noindent
Self-dual Yang-Mills theory in 2+2 dimensions arises as the target-space
dynamics of open strings with $N{=}2$ worldsheet supersymmetry,
whose topological nature renders the dynamics integrable~\cite{OV}.
By dimensional reduction one arrives at a Bogomolny system for Yang-Mills-Higgs
in 2+1 dimensions, which can be gauge fixed to the modified U($n$) chiral
model known as the Ward model~\cite{Ward88}. This model, though not Lorentz
invariant, is a rich testing ground for exact multi-soliton and wave 
configurations which, upon dimensional and algebraic reduction, descend
to multi-solitons of various integrable systems in 2+0 and 1+1 dimensions, 
such as sine-Gordon.

The two most popular deformations, noncommutativity and supersymmetry,
both preserve this integrability. Moyal-deformed extensions of the
above theories and their (mostly novel) solitonic solutions have recently
been studied in some detail~\cite{LPS2,LP1,LP2,B,W1,IU,Penati,DLP,ChuLe,KLP}. 
The goal of this talk is to add supersymmetry
to the game. Yang-Mills theory in 2+2 dimensions admits at most $\Ncal{=}4$
supersymmetries (16 supercharges), which limits the number of supersymmetries 
in the 2+1 dimensional Yang-Mills-Higgs system to $2\Ncal{=}8$. The Ward sigma
model based on the self-dual restriction inherits these supersymmetries.
These $2\Ncal{\le}8$ supersymmetric Ward models are somewhat non-standard, 
because their R-symmetry is non-compact, and their target space U($n$) is 
not constrained by the presence of supersymmetry~\cite{sigma8}. 

In this talk, which is based on~\cite{LP4}, I will consider
the supersymmetric noncommutative U($n$) Ward model and its multi-solitons,
with up to 16 supercharges and a Moyal deformation only of the two bosonic
spatial coordinates. To this end, I shall remind you of $\Ncal$-extended
self-dual Yang-Mills theory in 2+2 dimensions, the related super Bogomolny
system in 2+1 dimensions and the corresponding Ward model, before implementing
the standard Moyal deformation. In the operator formulation, the dressing
method will be employed to derive (second-stage) BPS conditions and to solve 
them, finally constructing U($n$) single- and multi-solitons including the
abelian case of~U(1). I will end with an outlook on current and future work.

\bigskip

\section{$\Ncal$-extended SDYM theory in 2+2 dimensions} \noindent
I begin with the four-dimensional Kleinian space $\R^{2,2}$ with
coordinates
\be
(x^\mu ) \= (x^a , \tilde t) \=(t,x,y,\tilde t) 
\qquad\text{where}\quad \mu=1,2,3,4 \quad\text{and}\quad a=1,2,3\ .
\ee
The signature $(-++\,-)$ allows me to introduce real isotropic coordinates
\be
x^{1\1} = \sfrac12(t-y)\ ,\quad x^{1\2} = \sfrac12(x+\tilde t)\ ,\quad
x^{2\1} = \sfrac12(x-\tilde t)\ ,\quad x^{2\2} = \sfrac12(t + y)
\ee
labelled by spinor indices $\a=1,2$ and $\da=\1,\2$.

The self-duality equations for a $u(n)$-valued field strength $(F_{\mu\nu})$
on $\R^{2,2}$ read
\be
\sfrac12\ve_{\mu\nu\r\s} F^{\r\s}=F_{\m\n}
\qquad\Leftrightarrow\qquad
F_{\da\db}\=0 \qquad\Leftrightarrow\qquad
F_{\a\da ,\b\db} \= \ve_{\da\db}\,F_{\a\b}
\ee
and are first-order differential equations for the gauge potentials
$A_{\a\da}\in u(n)$.

Let me add supersymmetry.
For $\Ncal\le4$, the field content of $\Ncal$-extended SDYM consists of
\be
A_{\a\da}\to F_{\a\b}\ ,\quad \chi^{i}_{\a}\ ,\quad \phi^{[ij]}\ ,\quad
\tilde\chi^{[ijk]}_{\da}\ ,\quad G_{\da\db}^{[ijkl]}
\qquad\text{with}\quad i,j,k,l=1,\ldots,\Ncal \ ,
\ee
where fields with an even (odd) number of spinor indices (anti)commute.
Their helicities range from $+1$ to $-1$. I refrain from writing the
field equations here.
It is convenient to introduce the $\Ncal$-extended superspace
$\R^{2,2|4\Ncal} \ni (x^{\a\da},\ \h_i^{\da},\ \th^{i\a})$
with derivatives
\bea
&\pa_{\a\da}\ :=\ \frac{\pa}{\pa x^{\a\da}}\ ,\qquad
\pa_{i\a}\ :=\ \frac{\pa}{\pa \th^{i\a}} \und
\pa^i_{\da}\ :=\ \frac{\pa}{\pa \h^{\da}_i}\ ,\\[4pt]
&D_{i\a} \= \pa_{i\a}+\h_i^{\da}\pa_{\a\da} \und
D^{i}_{\da} \= \pa^i_{\da}+\th^{i\a}\pa_{\a\da} \ .
\eea

An important subspace of $\R^{2,2|4\Ncal}$ is the {\sl chiral\/} superspace
$\R^{2,2|2\Ncal} \ni (x^{\a\da}{-}\th^{i\a}\h_i^{\da},\ \h_i^{\da})$.
It is relevant because the $\Ncal$-extended SDYM equations can be
rewritten in terms of superfields $\Acal_{\a\da}$ and $\Acal^i_{\da}$ 
on $\R^{2,2|2\Ncal}$. These chiral superfield potentials give rise to
chiral superfield strengths, whose leading components are the nonnegative
helicity fields:
\be
(\Acal_{\a\da},\ \Acal^i_{\da})\quad\to\quad
(\Fcal_{\a\b},\ \Fcal^i_\a,\ \Fcal^{ij})\ \supset\
(F_{\a\b},\ \chi^i_\a,\ \p^{ij}) \ .
\ee
Note that I employ calligraphic letters for chiral superfields.
With the help of chiral superspace covariant derivatives
\be
\nabla_{\alpha\da}\ :=\ \pa_{\alpha\da} +\Acal_{\alpha\da} \und
\nabla_{\da}^i\ :=\ \pa_{\da}^i+\Acal_{\da}^i
\ee
I can formulate $\Ncal$-extended self-duality as follows,
\be
[\nabla_{\a\da},\nabla_{\b\db}]\=\ve_{\da\db}\,\Fcal_{\a\b}\ ,\qquad
[\nabla^i_{\da},\nabla_{\b\db}]\=\ve_{\da\db}\,\Fcal^i_{\b}\ ,\qquad
\{\nabla^i_{\da},\nabla^j_{\db}\}\=\ve_{\da\db}\,\Fcal^{ij} \ .
\ee

These first-order equations may be viewed as the compatibility conditions of 
a linear system for a GL($n,\C$)-valued superfield $\psi(\z)$, namely
\be \label{linsys4}
\z^{\da}(\pa_{\a\da}+\Acal_{\a\da})\,\psi(\z) \= 0 \und
\z^{\da}(\pa_{\da}^i+\Acal_{\da}^i)\,\psi(\z) \= 0 
\ee
with a spectral parameter $\z\in\C P^1$ tucked into the spinor \ 
$\z_{\db}=\bigl(\begin{smallmatrix}1\\ \z\end{smallmatrix}\bigr)$
\ or \ $\z^{\da}=\ve^{\da\db}\z_{\db}=
\bigl(\begin{smallmatrix}\z\\ -1\end{smallmatrix}\bigr)$.

\bigskip

\section{$2\Ncal$-extended Bogomolny system in 2+1 dimensions} \noindent
Since the solitons I'd like to construct shall roam the noncommutative plane,
I may dimensionally reduce the $\Ncal$-extended SDYM system to a
$2\Ncal$-extended Bogomolny system one dimension lower. In chiral superspace,
this amounts to a reduction of 
$\ \R^{2,2|2\Ncal}\ $ to $\ \R^{2,1|2\Ncal}\ $
by demanding $\ \pa_{\tilde t}=0$ on all (super)fields. Furthermore, 
in 2+1 dimensions we may identify dotted with undotted spinor indices
and replace $\ \da,\db \to \a,\b$. It is well known that the number of
supersymmetries doubles in the process, i.e.\ $\Ncal\to2\Ncal$.
Therefore, we find a maximum of 8 supersymmetries in this 2+1 dimensional
Yang-Mills-Higgs system.

To be more explicit, I split the four coordinates into
\be
x^{[\a\b]} =  \sfrac12\ve^{\a\b}\tilde t \quad\text{and}\quad
x^{(\a\b)} =: y^{\a\b} \quad\text{with}\quad
y^{11}=\sfrac12(t{-}y)\ ,\quad
y^{12}=\sfrac12 x\ ,\quad
y^{22}=\sfrac12(t{+}y) 
\ee
and rewrite the spinorial derivations as
\be
D_{i\a} \=\pa_{i\a}+\h_i^{\b}\pa_{(\a\b )} \und
D_{\a}^i\=\pa_{\a}^i+\th^{i\b}\pa_{(\a\b )} \ .
\ee
Likewise, the $u(n)$ gauge potential decomposes as
\be
A_{\a\b} \= A_{(\a\b )} + A_{[\a\b ]} \=
A_{(\a\b )}-\ve_{\a\b}\,\vp \ ,
\ee
introducing the Higgs field $\vp$ as the fourth component,
and the covariant derivatives become
\be
D_{\a\b}\ :=\ D_{(\a\b)} \= \pa_{(\a\b)}\ +\ [A_{(\a\b)},\cdot\ ] \und
D_{[\a\b]} \= -\ve_{\a\b} [\vp,\cdot\ ]\ .
\ee
The corresponding field strength
\be
F_{\a\b,\,\g\de} \= [D_{\a\b},D_{\g\de}] \=
\ve_{\a\g}\,f_{\b\de}\ +\ \ve_{\b\de}\,f_{\a\g}
\qquad\text{with}\quad f_{\a\b}=f_{\b\a}
\ee
is entirely contained in the three components of $f_{\a\b}$
but subject to the Bogomolny equation
\be
f_{\a\b}\ +\ D_{\a\b}\vp \=0 \ ,
\ee
which implies the gauge and Higgs field equations of motion.

The $2\Ncal{=}8$ supersymmetric extension of the above yields 
the dimensional reduction of the $\Ncal{=}4$ SDYM equations.
The coupled equations for the multiplet 
$(F_{\a\b},\chi^i_\a,\p^{[ij]},\tilde{\chi}_{\a}^{[ijk]},G_{\a\b}^{[ijkl]})$
are
\bea
&&f_{\a\b}+D_{\a\b}\vp \=0\ ,
\non \\[4pt]
&&D_{\a\b}\,\chi^{i\b} + \ve_{\a\b}\,[\vp ,\, \chi^{i\b} ]\=0\ ,
\non \\[4pt]
&&D_{\a\b}\,D^{\a\b}\phi^{ij} + 2[\vp ,\, [\vp,\phi^{ij}]]+
2\{\chi^{i\a},\,\chi^j_{\a}\}\=0\ ,
\\[4pt]
&&D_{\a\b}\,\tilde{\chi}^{\b[ijk]}-\ve_{\a\b}\,[\vp,\,
\tilde{\chi}^{\b[ijk]} ]-6[\chi_{\a}^{[i},\ \phi^{jk]} ]\=0\ ,
\non \\[4pt]
&&D_{\a}^{\ \g} G_{\g\b}^{[ijkl]}+
[\vp,G_{\a\b}^{[ijkl]}]+
12\{\chi_{\a}^{[i},\tilde{\chi}_{\b}^{jkl]}\}
-18\,[\phi^{[ij},D_{\a\b}\phi^{{kl}]}]-
18\ve_{\a\b} [\phi^{[ij},[\phi^{{kl}]},\vp ]]\=0 \ .\phantom{XX}
\non
\eea

\bigskip

\section{$2\Ncal$-extended U($n$) Ward model in 2+1 dimensions} \noindent
Again, it is convenient to pass to chiral superfields
\be
(\Acal_{(\a\b)},\ \Xi,\ \Acal_\a^i)\ \supset\ (A_{(\a\b)},\ \vp,\ \ldots)\ ,
\ee
which are functions of $y^{\a\b}{-}\th^{i\a}\h_i^\b$ and $\h_i^\a$
and whose leading component in the $\h_i^\a$ expansion is indicated.
Predictably, the linear system (\ref{linsys4}) dimensionally reduces to
a linear system on $\R^{2,1|2\Ncal}$ of the form
\be \label{linsys3}
\z^{\b}(\pa_{(\a\b)} + \Acal_{(\a\b)}- \ve_{\a\b}\Xi)\,\psi \=0 \und
\z^{\a}(\pa_{\a}^i + \Acal_{\a}^i)\,\psi \=0 \ .
\ee

The gauge freedom can be used to fix the asymptotic form
of the GL($n,\C$)-valued auxiliary chiral superfield~$\psi$,
\be
\j(\z) \= \begin{cases}
{}\quad \Phi^{-1}\ +\ O(\z ) & \text{for}\quad\z\to 0 \\[8pt]
{}\quad \unity\ +\ \z^{-1}{\Ups}\ +\ O(\z^{-2} ) & \text{for}\quad\z\to\infty 
\end{cases} 
\ee
defining the Yang-type and Leznov-type prepotentials $\Phi$ and $\Ups$,
respectively. Multiplying (\ref{linsys3}) with $\psi^{-1}$ and
recalling that \
$\z^{\a}=\bigl(\begin{smallmatrix}\z\\ -1\end{smallmatrix}\bigr)$,
the asymptotics implies that
\bea
\Acal_{(21)}-\Xi&=&0 \und\qquad\!
\Acal_{(22)}\,\=\Phi^{-1}\pa_{t+y}\Phi\=\, \pa_{x}\Ups
\ =:\ \Acal \\[4pt]
\Acal_{(11)}&=&0 \und \
\Acal_{(12)}+\Xi\=\Phi^{-1}\pa_{x}\Phi\=\, \pa_{t-y}\Ups
\ =:\ \Bcal \\[4pt]
\Acal^i_1&=&0 \und\qquad\quad
\Acal^i_2\=\Phi^{-1}\pa^i_2\Phi\=\quad\pa^i_1\Ups
\ =:\ \Ccal^i
\eea
and determines $2{+}\Ncal$ nonvanishing potentials $\Acal$, $\Bcal$ and 
$\Ccal^i$ through either one of the two prepotentials. With this notation, 
the gauge-fixed linear system~(\ref{linsys3}) is spelled out as
\be \label{linsysfix}
(\z\pa_x - \pa_{t+y} - \Acal )\,\j \= 0\ ,\qquad
(\z\pa_{t-y} - \pa_x - {\cal B})\,\j \= 0\ ,\qquad
(\z\pa_{1}^i - \pa_{2}^i - {\cal C}^i )\,\j \= 0 \ .
\ee

This linear system's compatibility yields the 
$2\Ncal{\le8}$ Bogomolny equations in superspace form, 
which at the same time are equations of motion for the two prepotentials.
For the Yang-type prepotential $\Phi\in\tU(n)$ I obtain
\bea \label{Yanga}
\pa_x(\Phi^{-1}\pa_x\Phi) - \pa_{t-y}(\Phi^{-1}\pa_{t+y}\Phi)&=&0\ ,\\[4pt]
\pa_1^i(\Phi^{-1}\pa_x\Phi) - \pa_{t-y}(\Phi^{-1}\pa_2^i\Phi)&=&0\ ,\\[4pt]
\pa_1^i(\Phi^{-1}\pa_{t+y}\Phi) - \pa_x(\Phi^{-1}\pa_2^i\Phi)&=&0\ ,\\[4pt]
\pa_1^i(\Phi^{-1}\pa_2^j\Phi) + \pa_1^j(\Phi^{-1}\pa_2^i\Phi)&=&0\ ,
\label{Yangd}
\eea
which describes a supersymmetric extension of the Ward model --
an integrable chiral sigma model with WZW-like term --
on $\R^{2,1|2\Ncal}$. It is remarkable that this model enjoys up to
eight supersymmetries without any condition on its target space~\cite{sigma8}.
Alternatively, the equations of motion for the
Leznov-type prepotential $\Ups\in u(n)$ read
\bea \label{Leznova}
(\pa_x^2-\pa_{t+y}\pa_{t-y})\Ups+[\pa_{t-y}\Ups\,,\,\pa_x\Ups]&=&0\ ,\\[4pt]
(\pa_2^i\pa_{t-y}-\pa^i_1\pa_x)\Ups+[\pa^i_1\Ups\,,\,\pa_{t-y}\Ups]&=&0\ ,
\\[4pt]
(\pa_2^i\pa_x-\pa^i_1\pa_{t+y})\Ups+[\pa^i_1\Ups\,,\,\pa_x\Ups]&=&0\ ,\\[4pt]
(\pa_2^i\pa_1^j+\pa_2^j\pa^i_1)\Ups + \{\pa^i_1\Ups\,,\,\pa_1^j\Ups\}&=&0
\label{Leznovd}
\eea
and are merely quadratic.

\bigskip

\section{Moyal deformation} \noindent
In the remainder of this talk I shall construct solitonic solutions to
the supersymmetric Ward model equations~(\ref{Yanga})--(\ref{Yangd}). 
The integrability
of this model ensures the existence of multi-soliton configurations,
which can be found employing, e.g., the dressing method. To widen the scope,
I'd like to go one step further and subject the whole system to a
noncommutative deformation. It is known that the (nonsupersymmetric) Ward 
model can be Moyal-deformed without losing its integrability~\cite{LP1,LP2}.
What is more, the deformation enormously enhances the spectrum of solitons.
In particular, it allows for the novel class of abelian solitons, which
exist even in the U(1) case. It is to be expected that the supersymmetric
extension is compatible with this situation, so that the deformed 
supersymmetric Ward solitons are, one the one hand, superextensions
of the known bosonic noncommutative Ward solitons and, on the other hand,
deformations of (possibly singular) commutative super-Ward solitons.
In order to capture this wider class of solitons, I introduce at this stage 
a purely spatial Moyal deformation of $\R^{2,1}$ to $\R^{2,1}_\Th$ with
noncommutativity parameter $\Th>0$. Commutative spacetime can always be
recovered by taking the limit $\Th\to0$.

The deformation is effected by introducing the Moyal star product
\be
(f\sta g)(x,\th,\h) \= f(x^c,\th,\h) \exp\{\sfrac{\im}{2}\
\overleftarrow{\pa_a}\,\Th^{ab}\,\overrightarrow{\pa_b}\}\,g(x,\th,\h)
\ee
for functions on $\R^{2,1|4\Ncal}$, which for the coordinates yields
\be
x^a\sta x^b - x^b\sta x^a \=\im\Th^{ab}
\ee
with all other star (anti)commutators vanishing. In particular, I choose
not to deform the Grassmann coordinates to form a Clifford algebra, although
this could easily be implemented. The noncommutativity is parametrized by
the constant matrix $(\Th^{ab})$ whose entries I take to be
\be
\Th^{xy}=-\Th^{yx}=:\Th >0 \qquad\text{but}\qquad \Th^{tx}=\Th^{ty}=0\ .
\ee
Hence, $t$ as well as $\th^{i\a}$ and $\h_i^\a$ remain (anti)commutative.
For the complex coordinate combinations $z:=x+\im y$ and $\bar z:=x-\im y$
this reads
\be
\Th^{z\bar z} \= - \Th^{\bar z z} \= -2\im\,\Th \qquad\Rightarrow\qquad
z\sta\bar z - \bar z\sta z \= 2\,\Th\ .
\ee
Important properties of the star product are
\be
(f\,\sta\,g)\,\sta\,h \= f\,\sta\,(g\,\sta\,h) \und
\smallint\!\diff^2{z}\ f\sta g \= \smallint\!\diff^2{z}\, f\,g \ .
\ee

Instead of deforming the product in function space, one may alternatively
replace functions by operators, which act on an auxiliary Fock space~$\Hcal$.
To the noncommuting coordinates $z$ and $\bar z$ then correspond basic
operators $a$ and $a^\+$ subject to the Heisenberg-algebra relation
\be
[a\,,a^\+]\=\unity \ .
\ee
Their action on an orthonormal basis of ket states 
$\{|\ell\>,\ \ell\in\NN_0\}$ reads
\be
a\,|\ell\> \= \sqrt{\ell}\;|\ell{-}1\> \und
a^\+|\ell\> \= \sqrt{\ell{+}1}\;|\ell{+}1\> \ ,
\ee
which qualifies $|0\>$ as the vacuum state.

Let me restrict myself to the chiral superspace $\R^{2,1|2\Ncal}$.
The correspondence between functions $f(t,z,\bar z,\h^{\a}_i)$ and
operators $\hat f(t,\h^{\a}_i)$ is made precise by the
Moyal-Weyl map and its inverse,
\bea
f(t,z,\bar z,\h^{\a}_i) &\mapsto& \hat f(t,\h^{\a}_i)\=
\text{Weyl-ordered}\ f\bigl(t,\sqrt{2\Th}a,\sqrt{2\Th}a^\+,\h^{\a}_i\bigr)\ ,
\\[4pt]
F_{\sta} 
\bigl(t,\sfrac{z}{\sqrt{2\Th}},\sfrac{\bar z}{\sqrt{2\Th}},\h^{\a}_i\bigr)
&\mapsfrom& \hat F(t,\h^{\a}_i) \= F(t,a,a^\+,\h^{\a}_i)\ .
\eea
The crucial properties of this map are
\be
f\sta g \quad\mapsto\quad \hat f\ \hat g \und
{\textstyle\int}\diff x\ \diff y\ f \= 2\pi\,\Th\,\text{Tr}_\Hcal \hat f\ .
\ee
On the level of the coordinates, one has
\be
[t,\hat x]\=[t,\hat y]\=0 \quad\text{but}\quad [\hat x,\hat y]\=\im\Th
\qquad\Rightarrow\qquad [\hat z,\hat{\bar z}]\=2\,\Th\ ,
\ee
which implies
\be
\hat z \= \sqrt{2\Th}\,a \und
\pa_{\bar z}f \quad\mapsto\quad \hat\pa_{\bar z}\hat f \=
\sfrac{1}{\sqrt{2\Th}}\,[a\,,\hat f] \ .
\ee

\bigskip

\section{Dressing method} \noindent
In order to construct multi-soliton configurations, I shall employ
the dressing method to build up solutions to the gauge-fixed linear system
(\ref{linsysfix}). The latter, together with a gauge-compatible reality 
condition, may be written as
\be \label{real}
\j(x^a,\h_i^{\a},\z)\ \bigl[\j(x^a,\h_i^{\a},\bar\z)\bigr]^\+\=\unity \und
\ee
\be \label{linsys}
\j (\pa_{t+y} - \z\pa_x )\j^\+ \= \Acal \ ,\qquad
\j (\pa_x - \z\pa_{t-y} )\j^\+ \= \Bcal \ ,\qquad
\j (\pa_2^i - \z\pa_1^i )\j^\+ \= \Ccal^i\ .
\ee

The result of iterating the dressing procedure is a multi-pole ansatz
for the operator-valued $n{\times}n$ matrix $\j$,
\be \label{ansatz}
\j_m\=\prod\limits_{\ell=0}^{m-1}\Bigl(\unity\ +\
\frac{\mu_{m-\ell} -\bar\mu_{m-\ell}}{\z - \mu_{m-\ell}}\ P_{m-\ell}\Bigr)
\= \unity\ +\ \sum\limits_{k=1}^{m}\frac{\La_{mk}S^\+_k}{\z - \mu_{k}}\ ,
\ee
whose multiplicative and additive forms contain square matrices 
$P_{m-\ell}(x^a,\h_i^{\a})$ and rectangular matrices
$\La_{mk}(x^a,\h_i^{\a})$ and $S_k(x^a,\h_i^{\a})$, respectively,
which are to be determined. The complex parameters $\mu_1,\mu_2,\ldots,\mu_m$
tell the positions of the $\z$-poles of the meromorphic function~$\j$.
Such poles must occur for a nontrivial $\bar\z$-independent function 
on~$\C P^1$.

Observe now that the right-hand sides of (\ref{real}) and~(\ref{linsys})
are independent of $\z$, which implies that the residue of any $\z$-pole
on their right-hand sides must vanish. It suffices to consider the residues
at $\z{=}\bar\m_k$, which for the additive form of (\ref{ansatz}) yields
\be \label{residues}
\Bigl(\unity\ +\ \sum\limits^{m}_{\ell=1}
\frac{\La_{m\ell}S^\+_\ell}{\bar\mu_k - \mu_\ell}\Bigr)\,S_k\ \=0\=\
\Bigl(\unity\ +\ \sum\limits^{m}_{\ell=1}
\frac{\La_{m\ell}S^\+_\ell}{\bar\mu_k-\m_\ell}\Bigr)\,
\bar L_k^{\Acal,\Bcal,\Ccal}\,S_k \ ,
\ee
where $\bar L_k^{\Acal,\Bcal,\Ccal}$ stands for either
\be \label{Lbar}
\bar L_k^{\Acal} \= \pa_{t+y} -\bar\mu_k\pa_x\ ,\qquad
\bar L_k^{\Bcal} \= \mu_k(\pa_x -\bar\mu_k\pa_{t-y})
\qquad\text{or}\qquad
\bar L_k^{\Ccal} \= \pa^i_2 - \bar\mu_k\pa_1^i \ .
\ee

\bigskip

\section{Co-moving frames} \noindent
The left equation in (\ref{residues}) is algebraic and will later be used
to find $\La_{m\ell}$. Assuming its validity for the moment, the right
equation is satisfied if I demand that $S_k$ is an eigenobject of the
$2{+}\Ncal$ differential operators in~(\ref{Lbar}). This requirement
determines the dependence of~$S_k$ on $t$ and on one-half of the $\h_i^\a$.
To see that, it is useful to pass, seperately for each value of $k$, from 
the coordinates $(t,x,y;\h_i^1,\h_i^2)$ to (complex) ``co-moving coordinates''
$(w_k,\bar w_k,s_k;\h^i_k,\bar\h^i_k)$ via
\be
w_k\ :=\ \sfrac12(\bar\mu_k{+}\sfrac1{\bar\mu_k})t
+x+\sfrac12(\bar\mu_k{-}\sfrac1{\bar\mu_k})y \und
\h^i_k\ :=\ \h_i^1+\bar\mu_k\h_i^2 \ .
\ee
It is easy to check that the operator version of $w_k$ obeys
\be
[\hat w_k\,,\,\hat{\bar w}_k]\= 2\n_k\bar\n_k\,\Th
\qquad\text{with}\qquad
\nu_k\bar\nu_k \= \sfrac{4\im}{\mu_k-\bar\mu_k-\mu_k^{-1}+\bar\mu_k^{-1}}\cdot
\sfrac{\mu_k-\mu_k^{-1}-2\im}{\bar\mu_k-\bar\mu_k^{-1}+2\im} \ .
\ee
This suggests me to define co-moving annihilation and creation operators,
\be
\hat w_k\ =:\ \sqrt{2\Th}\,\n_k\,c_k \und
\hat{\bar w}_k\ =:\ \sqrt{2\Th}\,\bar\n_k\,c_k^\+
\qquad\text{so that}\quad [c_k\,,\,c_k^\+]\=\unity \ .
\ee
These operators are all related to $a$ and $a^\+$ by inhomogeneous Bogoliubov
transformations via specific squeezing operators $U_k(t)\in \text{ISU(1,1)}$,
so that
\be \label{covac}
c_k\,|\vel_k\>\=0 \qquad\text{defines a co-moving vacuum}\quad
|\vel_k\> \= U_k(t)|0\> \ .
\ee
For a given value of $k$, the co-moving coordinates define a ``rest frame''
through the condition $\frac{\pa}{\pa s_k}\=0$, which linearly relates
the coordinates $x$, $y$ and $t$, so that one gets a straight trajectory 
$(x(t),y(t))$ in the $xy$~plane. Since these trajectories turn out to describe
the motion of individual lumps of energy in a generic multi-soliton 
configuration, I can compute their various velocities,
\be
(\dot{x}(t)\,,\,\dot{y}(t))_k \ \equiv\
(\text{v}_x\,,\,\text{v}_y)_k \= - \Bigl(
\frac{\mu_k+\bar\mu_k}{\mu_k\bar\mu_k+1}\;,\,
\frac{\mu_k\bar\mu_k-1}{\mu_k\bar\mu_k+1} \Bigr)\ ,
\ee
as a function of the corresponding pole positions.
Clearly, there is no scattering. However, when some of these velocities 
coincide, more general time dependence (including scattering) arises.
The relation between $\vec{\text{v}}$ and $\mu$ is depicted in Figure~1,
and Figure~2 sketches the worldline of a single soliton in the two
coordinate frames. 
\begin{figure}
\begin{center}
\includegraphics[width=10cm]{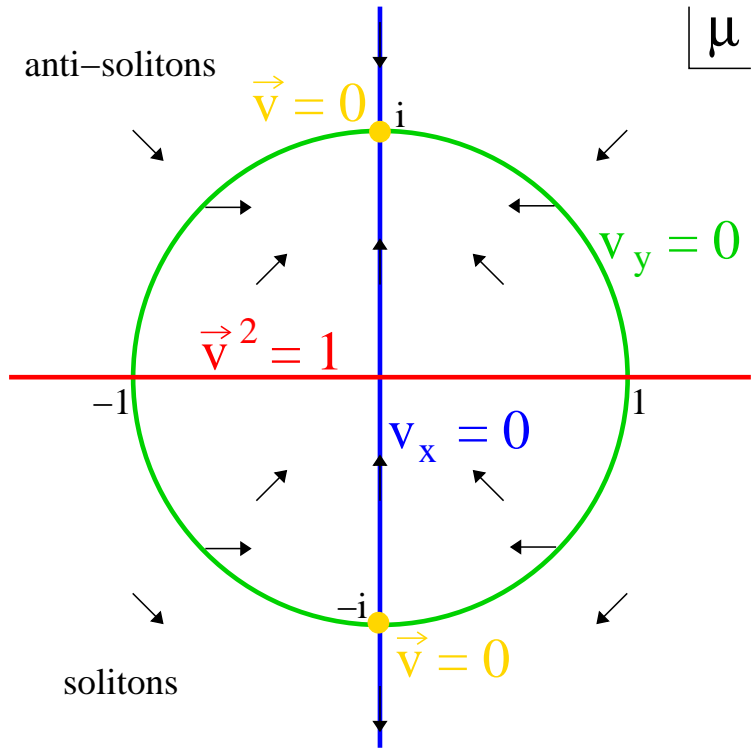}
\end{center}
\caption{Soliton velocity $\vec\vel$ as a function of $\mu$}
\end{figure}
\begin{figure}
\begin{center}
\includegraphics[width=10cm]{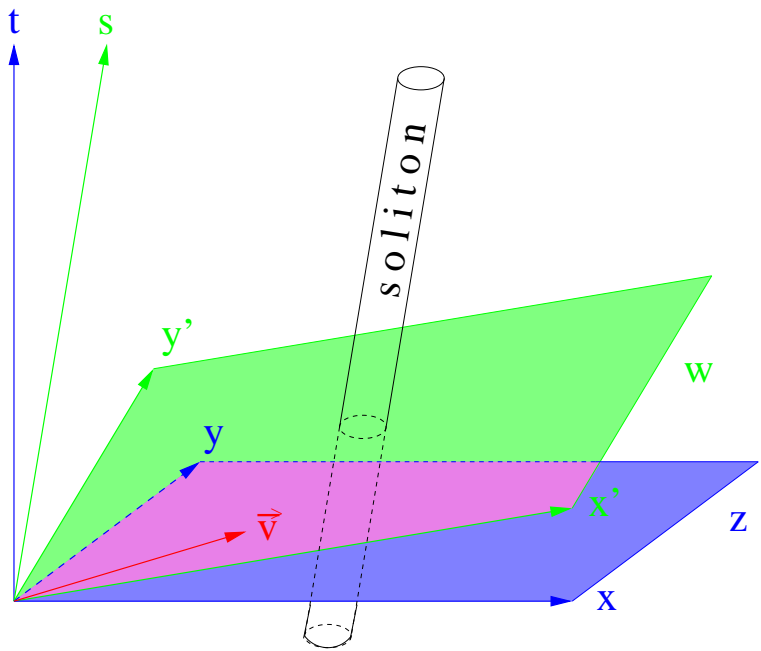}
\end{center}
\caption{Soliton worldline in reference and rest frames}
\end{figure}

\section{BPS conditions} \noindent
Recall the vanishing-residue conditions (\ref{residues}),
\be \label{residue2}
\bigl(\unity\ +\ \Sigma_k\bigr)\,S_k \= 0\=
\bigl(\unity\ +\ \Sigma_k\bigr)\,\bar L_k^{\Acal,\Bcal,\Ccal}\,S_k\ ,
\ee
where $\Sigma_k$ abbreviates the sum over~$\ell$. 
The left equation, which arose from the reality condition~(\ref{real}), 
will be solved a bit later, using the multiplicative form of the 
ansatz~(\ref{ansatz}). For the right equation, which represents the linear
system~(\ref{linsys}) and therefore the equation of motion, the differential
operators simplify in the co-moving coordinates to
\be
\bar L_k^{\Acal} \= \bar L_k^\Bcal \=
(\mu_k{-}\bar\mu_k)\frac{\pa}{\pa\bar w_k}\ =:\ \bar L_k \und
\bar L_k^{\Ccal} \= (\mu_k{-}\bar\mu_k)\frac{\pa}{\pa\bar \h_k^i}\ =:\
\bar L_k^i\ .
\ee
Essentially, this means that $S_k$ does not depend on $\bar w_k$ or 
$\bar \h_k^i$. However, since $(\unity+\Sigma_k)$ is not invertible,
in view of the left equation in (\ref{residue2}) slightly weaker conditions 
hold, namely
\be
\bar L_k\,S_k \= S_k\,\tilde Z_k \und\,
\bar L_k^{i}\,S_k \= S_k\,\tilde Z_k^i \ .
\ee
In the operator formulation of the Moyal deformation, these read\footnote{
The matrices $\hat S_k$ being rectangular, left multiplication by $c_k$ is
the natural operator equivalent to the $\bar L_k$ action. However, one may
instead use the adjoint $c_k$ action by redefining $\hat Z_k$.}
\be
c_k\,\hat S_k \= \hat S_k\,\hat Z_k \und
\sfrac{\pa}{\pa\bar\h_k^i}\,\hat S_k \= \hat S_k\,\hat Z_k^i\ .
\ee
To unclutter the formulae I drop the hats and temporarily suppress the
label~$k$. The noncommutative BPS equations then takes the form
\be \label{ncBPS}
c\,S \= S\,Z \und \sfrac{\pa}{\pa\bar\h^i}\,S \= S\,Z^i\ ,
\ee
where solutions $S_k$ are parametrized by $Z$ and $Z^i$.
 
The goal is to find operator-valued $n{\times}r$ matrices
$S(c,c^\+,\h^j,\bar\h^j)$ for any choice of $r{\times}r$ matrices 
$Z^{(i)}(c,c^\+,\h^j,\bar\h^j)$.
Recall that $n$ is determined by the group U($n$) but $r$ is arbitrary
and in general depends on~$k$.
The general solution to (\ref{ncBPS}) is rather involved, but things
simplify when I restrict myself to holomorphic parameter matrices
\be
Z\=Z(c,\h^j) \und Z^i\=Z^i(c,\h^j) \ .
\ee
 
\bigskip

\section{Nonabelian BPS solutions} \noindent
For holomorphic choices of $Z$ and $Z^i$ I can write down the general solution
to (\ref{ncBPS}) as
\be \label{BPSsol}
S \= R(c,\h^i)\
:\!\exp\bigl\{Z(c,\h^j)\,c^\+ - c\,c^\+\bigr\}\!:\
\exp\bigl\{\textstyle{\sum_i} Z^i(c,\h^j)\,\bar\h^i\bigr\}\
Q(c,\h^j) \ ,
\ee
where $R_{n\times r}$ and $Q_{r\times r}$ are arbitrary
holomorphic matrix functions of $c$ and $\h^j$ and where the colons
signify normal ordering with respect to the $(c,c^\+)$ algebra.
It turns out that 
\be
S\quad\to\quad S\,\Ga \und Z^{(i)}\quad\to\quad \Ga^{-1} Z^{(i)}\,\Ga
\ee
does not change the final solution $\Phi$ or $\Ups$ and, hence,
invertible right factors in the solution (\ref{BPSsol}) may be omitted.
Without loss of generality, I therefore put $Z^i=0$.
A regular matrix~$Q$ allows me to simplify the solution (\ref{BPSsol}) to
\be \label{BPSsol2}
S \= R(c,\h^i)\
:\!\exp\bigl\{Z(c,\h^j)\,c^\+ - c\,c^\+\bigr\}\!:
\ee
and further to
\be \label{BPSsol3}
S \= R(c,\h^i)
\ee
if the normal-ordered exponential is invertible.
 
The prime example for the latter, holomorphic, case is 
\be
Z=\unity_r\otimes c \qquad\Leftrightarrow\qquad
[c\,,\,S] \= 0 \ ,
\ee
which immediately sets $S=R$. For $r{<}n$ I can interpret $S$ as a map
\be
S:\ \C^r\otimes\Hcal\ \hookrightarrow\ \C^n\otimes\Hcal\ ,
\ee
and (\ref{ncBPS}) says that this embedding is stable under the action of $c$ 
and of $\frac{\pa}{\pa\bar\h^i}$. The ensueing solitons I call ``nonabelian''
because they require $n\ge2$ and in the commutative ($\Th{\to}0$) limit
smoothly reproduce the undeformed (supersymmetric) Ward-model solitons.
 
\bigskip

\section{Abelian BPS solutions} \noindent
The situation is different in the (more general) case where
$:\!\e^{-c^\+(c-Z)}\!:$ has a nonzero kernel. In particular, this occurs
whenever the $r{\times}r$ matrix~$Z$ is not operator-valued, i.e.
\be
Z=Z(\h^j) \qquad\Rightarrow\qquad
:\!\e^{-c^\+(c-Z)}\!: \= \e^{Zc^\+} :\!\e^{-c^\+c}\!: \=
\e^{Zc^\+}\,|\vel\>\<\vel| \ =:\ |Z\>\<\vel| \ ,
\ee
where I defined an $r{\times}r$ matrix $|Z\>$ of (unnormalized) coherent states
(eigenstates of~$c$), whose parameters are still functions of the~$\h^j$. 
Recalling the freedom of right multiplication with~$Q$, 
the resulting solution reads
\be
S\= R(c,\h^i)\,:\!\e^{-c^\+(c-Z(\h^j))}\!:\,Q(c,\h^i) 
\= R(Z,\h^j)\ |Z\>\<*|
\ee
with an arbitrary $r{\times}r$ matrix~$\<*|$ of bras.
 
Clearly, the factor of $|Z\>\<*|$ may not be dropped here.
However, I can strip off the $\<*|$~piece by generalizing $S$
to {\it any\/} $c$-stable embedding $Y\hookrightarrow\C^n\otimes\Hcal$.
For $\textrm{dim}Y\equiv q<\infty$ an embedding is just 
an $n{\times}q$ array of kets,
\be
|S\>:\ \C^q\ \hookrightarrow\ \C^n\otimes\Hcal\ ,
\ee
and the BPS conditions~(\ref{ncBPS}) become
\be \label{ncBPSket}
c\,|S\> \= |S\>\,Z(\h^j) \und
\sfrac{\pa}{\pa\bar\h^i}\,|S\> \= |S\>\,Z^i(\h^j) \ .
\ee
Please note that $n=1$ is now possible, i.e.\ there exist novel solitons 
in the noncommutative U(1)~model!
Keeping $Z^i=0$ and removing $Q(\h^i)$, the BPS solution takes the form
\be
|S\> \= R(c,\h^i)\ \e^{Z(\h^j)c^\+}\,|\vel\> \= R(Z,\h^j)\ |Z\> \ .
\ee
Generically, I can employ the right multiplication with $Q$ to diagonalize
\be
Z=\text{diag}\,(\a^1,\ldots,\a^q) \qquad\Rightarrow\qquad
|Z\> \= \text{diag}\,\bigl(|\a^1\>,\ldots,|\a^q\>\bigr)
\qquad\text{with}\quad |\a^l\>:=\e^{\a^l(\h^j) c^\+}|\vel\> \ .
\ee
{}From the Moyal-Weyl correspondence
$|\vel\>\<\vel|\ \leftrightarrow\ 2\,\e^{-w\bar w/\Th}$
it is obvious that these solutions become singular in the commutative 
($\Th{\to}0$) limit. Therefore, I call them ``abelian solutions''.
 
\bigskip

\section{One-soliton configuration} \noindent
I now build up the full soliton comfigurations from the building blocks
$S_k$ or $|S_k\>$. Let me first present the simplest case, namely $m=1$,
and suppress the index~$k$. The ansatz~(\ref{ansatz}) reduces to
\be
\psi_1 \ \=\ \unity\ +\ \frac{\mu -\bar\mu}{\z -\mu}\,P
\ \=\ \unity\ +\ \frac{\La\ S^\+}{\z -\mu}\ .
\ee
This is a good moment to work out the reality condition~(\ref{real})
in terms of the operator-valued $n{\times}n$ matrix~$P$.
The vanishing of the residue at $\z{=}\bar\mu$ immediately yields
\be
P \= P^\+ \= P^2 
\qquad\Rightarrow\qquad
P\=T\,\frac1{T^\+ T}\,T^\+ \quad\text{or}\quad
\= |T\>\,\frac{1}{\<T|T\>}\,\<T| \ .
\ee
Similarly, the linear system~(\ref{linsys}) is residue-free for
\bea
(\unity{-}P)\,c\,P \= 0 
\quad&\Rightarrow&\quad\quad 
c\,T \= T\,Z \ \qquad\text{or}\qquad\quad
c\,|T\> \= |T\>\,Z \ ,\\[4pt]
(\unity{-}P)\,\sfrac{\pa}{\pa\bar\h^i}\,P \= 0
\quad&\Rightarrow&\quad\,
\sfrac{\pa}{\pa\bar\h^i}\,T \= T\,Z^i \qquad\text{or}\qquad
\sfrac{\pa}{\pa\bar\h^i}\,|T\> \= |T\>\,Z^i \ .
\eea
 
Comparison with (\ref{ncBPS}) and (\ref{ncBPSket}) shows that $T=S$ 
and $\Lambda=(\mu{-}\bar\mu)T\frac1{T^\+T}$ in this situation.
Borrowing from the previous two sections, I generically have
\bea
T\ \=\ \,S\ &=& R(c,\h^j)_{n\times r} \ ,
\qquad\text{e.g. polynomials in $(c,\h^j)$}\ , \\[4pt]
|T\> \= |S\> &=& R(c,\h^j)_{n\times q}\
\biggl(\begin{smallmatrix}
|\a^1(\h^j)\> & & \\[-6pt] & \ddots & \\ & & |\a^q(\h^j)\>
\end{smallmatrix} \biggr)_{q\times q} \ .  \label{Tabelian}
\eea
I remark that each matrix element of $R$ enjoys an $\h$-expansion
\be \label{Rexp}
R(c,\h^i) \= R_0(c) + \h^i R_{[i]}(c) + \h^i\h^j R_{[ij]}(c) +
\h^i\h^j\h^k R_{[ijk]}(c) + \h^i\h^j\h^k\h^l R_{[ijkl]}(c)\ ,
\ee
and $c$ acting on a coherent state $|\a^l\>$ may be replaced by its
eigenvalue~$\a^l$. From the above expressions it is only a matter of
diligence to obtain the prepotentials via
\be
\Phi \= \unity\ -\ (1{-}\sfrac{\m}{\bar\m})P \und \Ups \= (\m{-}\bar\m)P
\ee
and hence any other quantity one desires. Finally I remark that
nonabelian and abelian solutions are distinguished by the rank of the
projector~$P$: in the former case $P$ has infinite rank in $\C^n\otimes\Hcal$
but finite matrix rank~$r$ in~$\C^n$, while in the latter case $P$ is of
finite rank~$q$ in~$\C^n\otimes\Hcal$.
 
\bigskip

\section{$\Ncal$-extended multi-soliton configurations} \noindent
The real benefit of integrability is the existence of multi-soliton solutions 
to our sigma-model equations~(\ref{Yanga})--(\ref{Yangd}) or quadratic 
equations~(\ref{Leznova})--(\ref{Leznovd}).
The dressing method provides me with an iterative scheme to build
increasingly complex classical configurations, adding one lump at a time.
First, the ansatz~(\ref{ansatz}) contains $m$ pole positions~$\mu_k$,
which we take from the lower complex plane to exclude anti-solitons.
Each $\mu_k$ gives rise to its own co-moving coordinate~$w_k$ and,
via Moyal deformation, to a Heisenberg pair~$(c_k,c_k^\+)$ with its own
vacuum~$|\vel_k\>$. Using the additive form of the ansatz, I have shown how 
to build nonabelian or abelian BPS solutions $S_k$ or $|S_k\>$ on this vacuum.
Second, the multiplicative form of the ansatz generates conditions on
the projectors~$P_k$ and thus on the corresponding $T_k$ or $|T_k\>$. 
These conditions easily allow one to relate $T_k$ with $S_k$ in general,
\be
T_1 \= S_1 \und
T_k \=\biggl\{\prod\limits_{\ell =1}^{k-1}\Bigl(\unity\ -\
\frac{\mu_{k-\ell} -\bar\mu_{k-\ell}}{\mu_{k-\ell} - \bar\mu_{k}}\
P_{k-\ell}\Bigr)\biggr\}\ S_k \qquad\text{for}\quad k\ge2\ ,
\ee
{}from which the projectors
\be
P_k \= T_k\,\frac{1}{T_k^\+T_k}\,T_k^\+ \qquad\text{or}\quad
P_k \= |T_k\>\,\frac{1}{\<T_k|T_k\>}\,\<T_k|
\ee
are readily expressed in terms of the matrices $R_k$ and $Z_k^{(i)}$, 
whose size may vary with~$k$. These matrices contain all moduli of the
multi-soliton configuration besides the ``rapidities''~$\mu_k$.
With the projectors in hand, I do not need the form of the $\Lambda_{mk}$,
and the prepotentials are known as
\be
\Phi \ = \prod_{\ell=0}^{m-1}
\bigl(\unity-(1{-}\sfrac{\m_{m-\ell}}{\bar\m_{m-\ell}})P_{m-\ell}\bigr)
\quad\ \text{and}\ \quad
\Ups \ = \sum_{k=1}^{m} (\mu_{k}{-}\bar\mu_{k}) P_k \ .
\ee
 
As the simplest example let me present the U(1) two-soliton
($n{=}1$, $m{=}2$) with ranks $q_1=q_2=1$.
For $n{=}1$, the row matrix~$R$ in~(\ref{Tabelian}) affects only the
normalization of the coherent states~$|\a^l\>$ and is thus irrelevant.
It remains the two rapidities $\mu_1$ and~$\mu_2$ as well as two complex
coherent-state parameters $\alpha_1$ and~$\alpha_2$, which may be functions
of the Grassmann coordinates~$\h^i$.
The result of the given recipe eventually produces the Yang-type prepotential
\be
\Phi^\+ \= \unity\ -\ \frac{1}{1-\m|\s|^2} \,\Bigl[\,
\frac{\m_{11}}{\m_1}|1\>\<1|\ +\
\frac{\m_{22}}{\m_2}|2\>\<2|\ -\
\s\m\frac{\m_{21}}{\m_2}|1\>\<2|\ -\
\bar\s\m\frac{\m_{12}}{\m_1}|2\>\<1| \,\Bigr] 
\ee
with  ($k=1,2$)
\be
|k\>\ \propto\ |\a_k\>\ ,\qquad \<k|k\>=1\ ,\qquad \<1|2\>\ =:\ \s\ ,\qquad
\m_{ij}\ :=\ \m_i{-}\bar\m_j\ ,\qquad
\m\ :=\ \frac{\m_{11}\,\m_{22}}{\m_{12}\,\m_{21}}\ ,
\ee
and the time dependence hides in the squeezing operators $U_k(t)$ that relate
the vacua $|\vel_k\>$ (on which the $|\a_k\>$ are built) to the reference 
vacuum~$|0\>$ (see~(\ref{covac})).
 
\bigskip

\section{Outlook} \noindent
I have arrived at explicit multi-soliton configurations on~$\R^{2,1|2\Ncal}$,
which formally look like the solutions in the non-supersymmetric model,
except for additional dependencies on the chiral Grassmann coordinates~$\h^i$
via the matrices $R_k(c_k,\h_k^j)$ and the parameters $\a_k^l(\h_k^j)$ 
entering $S_k$ and $|S_k\>$. Furthermore, the Grassmann-odd components in
any $\h$-expansion, such as~(\ref{Rexp}), require the introduction of 
extraneous Grassmann-odd moduli $\ve_i$, like in $R=R_0+\h^i\ve_i R'+\ldots$.

As a simple example, consider the static abelian rank-one soliton 
($\mu{=}-\im\leftrightarrow\vel{=}0$), based on a coherent state
\be
|T\> \= \e^{\a(\h^i)a^\+}\,|0\> \= 
\sum_{p=0}^4 \a_b^p (a^\+)^p |\a_s\>\ ,
\ee
with (subscripts referring to ``body'' and ``soul'')
\be
\a(\h^i) \= \a_b + \a_s \=
\a_b + \h^i \a_{[i]} + \h^i\h^j \a_{[ij]} +
\h^i\h^j\h^k \a_{[ijk]} + \h^i\h^j\h^k\h^l \a_{[ijkl]}\ .
\ee
Simplifying to $\a_b{=}0$ so that $\a^5=\a_s^5=0$, the projector becomes
\bea
P &=& \Bigl( {\textstyle\sum_{p=0}^4}\,p!(\bar\a_s\a_s)^p \Bigr)^{-1}\
\sum_{k,l=0}^4 \sfrac{\a_s^k\bar\a_s^l}{k!l!}\,|k\>\<l| \\[4pt]
&=& (1-\bar\a_s\a_s+\ldots)\,\bigl( 
|0\>\<0| + \a_s|1\>\<0| + \bar\a_s|0\>\<1| + \a_s\bar\a_s|1\>\<1| +\ldots\bigr)
\ .
\eea
For $\Ncal{=}1$, I must put $\a_s=\h\ve$ with $\ve^2=0$ and get
\be
P \= |0\>\<0|\ +\ \h\ve\,|1\>\<0|\ -\ \bar\h\bar\ve\,|0\>\<1|\ +\
\h\bar\h\ve\bar\ve\,\bigl(|1\>\<1|-|0\>\<0|\bigr)\ .
\ee
Employing the Moyal-Weyl map, this shows how the supersymmetric extension
modifies the profile of the basic soliton. I have also checked the simplest
U(2) examples, e.g.\ 
$T=\bigl(\begin{smallmatrix}\a(\h)\\a+\b(\h)\end{smallmatrix}\bigr)$, with
similar results. In all cases, the topological charge is determined by the
body component alone.

One would also like to investigate the energy density of such solitons.
However, beyond the static $\Ncal{=}1$ configurations, no action or energy
functional is known for our $\Ncal$-extended sigma models, so this remains 
a challenge.

How about scattering? To learn about the lump-lump interaction inside
a non-static multi-soliton solution, I focus on the large-time asymptotics.
It is easy to see that, in the $(t,x,y)$ frame, a moving lump yields
\be
T \quad\buildrel{{|t|\to\infty}}\over\longrightarrow\quad t^q\ \Gamma
\qquad\Rightarrow\qquad
P \quad\buildrel{{|t|\to\infty}}\over\longrightarrow\quad
\Pi \= \Gamma\,\sfrac1{\Gamma^\+\Gamma}\,\Gamma^\+
\ee
with a constant $n{\times}r$ matrix $\Gamma$,
so that
\be
\lim_{t\to\pm\infty}\Phi_m \= 
\bigl(\unity-(1{-}\sfrac{\mu_m}{\bar\mu_m})\Pi_m\bigr) \cdots
\bigl(\unity-(1{-}\sfrac{\mu_2}{\bar\mu_2})\Pi_2\bigr)
\bigl(\unity-(1{-}\sfrac{\mu_1}{\bar\mu_1})\Pi_1\bigr)\ .
\ee
If the $k$th lump is static, $\Pi_k$ must be replaced by~$P_k$.
The important message here is that 
\be
\lim_{t\to+\infty}\Phi_m\=\lim_{t\to-\infty}\Phi_m\ ,
\ee
which proves that the individual lumps in these multi-soliton
configurations do not feel each other at all, but keep moving
along their straight trajectories with constant velocities in
the Moyal plane. When two lumps' velocities coincide, however, 
the situation changes. In the limit of vanishing relative velocities
a new time dependence is known to emerge for the non-supersymmetric
Ward solitons, giving rise to breather-type and to nonabelian scattering 
configurations~\cite{LP2,KLP}. 
I expect this feature to carry over to the models considered here.

Finally, one may ask why I have deformed only the bosonic coordinates
but not also the fermionic ones. The reason was merely a practical one; 
the more general case of Moyal-deforming the whole $\R^{2,1|2\Ncal}$
superspace is an obvious next step. Of special interest may be the
other extreme, a purely fermionic deformation. It should lead to
non-{\it anti\/}commuting solitons based on the Clifford algebra
\be
\{\h_i^{\a}\,,\,\h_j^{\b}\}\=C_{ij}^{\a\b}\ .
\ee
I foresee the corresponding BPS solutions to break part of the supersymmetry
and perhaps carry additional structure interpretable as spin.

\bigskip

\ack \noindent
I thank C.~Gutschwager, T.A.~Ivanova and A.D.~Popov
for various fruitful discussions and collaborations.
This work was partially supported by the Deutsche Forschungsgemeinschaft.

\end{document}